\documentclass[fleqn,12pt,twoside]{article}
\usepackage{espcrc1}

\usepackage{graphicx}
\usepackage[figuresright]{rotating}

\usepackage{amsmath,amsfonts,epsf}
\usepackage{ascii}
\usepackage{xcolor}
\usepackage{soul}
\usepackage{ulem}
\usepackage{braket}  
\newcommand{\half}{\tfrac{1}{2}}

\newcommand{\be}{\begin{equation}}
\newcommand{\ee}{\end{equation}}
\newcommand{\beq}{\begin{equation}}
\newcommand{\eeq}{\end{equation}}
\newcommand{\bea}{\begin{eqnarray}}
\newcommand{\eea}{\end{eqnarray}}

\title{
Frank Discussion of the Status of Ground-state Orbital-free DFT
}

\author{Valentin V.~Karasiev
\address[QTP]{Quantum Theory Project, 
         Departments of Physics and of Chemistry,
         University of Florida, Gainesville FL 32611-8435}$~^*$,
S.B.~Trickey
\addressmark[QTP]
        \thanks{Work supported 
by U.S.\ Dept.\ of Energy grant $\rm DE$-$\rm SC$0002139.}%
}

\begin{document}


\maketitle

\tableofcontents

\begin{abstract}
\noindent
F.E.\ Harris has been a significant partner in our work
on orbital-free density functional approximations for use in
{\it ab initio} molecular dynamics.  Here we mention briefly the
essential progress on single-point functionals since our 
original paper (2006). Then we focus on the advantages and 
limitations of generalized gradient  approximation (GGA) 
non-interacting  kinetic-energy functionals.  We reconsider
the constraints provided by near-origin conditions in atomic-like
systems and their relationship to regularized versus physical 
external potentials. Then we seek the best empirical GGA for
the non-interacting KE for a modest-sized set of molecules
with a well-defined near-origin behavior of their 
densities.  The search is motivated by a desire for insight into 
GGA limitations and for a target for constraint-based development.
\end{abstract}



\section{Introduction} 
\subsection{Background}

At the Harris Workshop (10-12 Dec.\ 2014), the second author spoke
about recent progress by the Univ.\ Florida orbital-free density functional
theory (OFDFT) group of which Frank Harris is a member.  
Substantially all that work is reported in 
Refs.\ \cite{PRL112-076403,CPC185-3240,PRB88-161108R} and references
therein.  Earlier work and voluminous references for context are in
two review articles from our group \cite{PostNAMETOFKEreview,IPAMreview}.  \vspace*{-1pt}

Rather than recapitulate the talk and publications, here we 
provide a particular perspective on what has been learnt.  
The phrase ``near-origin''
rather than ``cusp condition'' is a clue to the role  the 
external potential plays in  enforcing behavior upon 
generalized gradient approximations (GGAs) for the non-interacting 
kinetic energy.    We present some new
results on near-origin conditions applied to GGAs.  These 
extend work we did with Frank Harris \cite{JCAMD,Signpost,KJTH-PRB80}.
Then we explore implications of a generic regularization of the 
usual external potential (from a nuclear array) by empirical determination
of the most nearly optimal GGA for a set of molecular data.  
That continues the study of binding in simple molecules by 
non-self-consistent OFDFT  with
key ingredients of the methodology introduced in our previous publications
\cite{JCAMD,Signpost,KJTH-PRB80}. These ingredients, besides the
near-origin analysis of the Pauli potential, include
(i) the use of a set of nuclear spatial configurations for the same molecule;
(ii) the use of Gaussian Kohn-Sham molecular densities as input;
(iii) so-called $\Delta E$ criterion which enforces binding and
(iv) the $E$ criterion which enforces correct absolute energies.
See also the recent work of Borgoo et al.\  \cite{Borgoo..Tozer.binding.2014}
in which the relationship between binding and the effective 
homogeneity of approximate non-interacting kinetic energy functional is 
considered. \vspace*{-1pt}

Though our research agenda emphasizes 
functionals for free 
energy DFT \cite{Mermin65} primarily for use in the 
warm dense matter regime, here we restrict attention to ground state 
OFDFT.  There are three reasons. First, ground state OFDFT is a hard challenge 
(as history going all the way back to Thomas \cite{Thomas}
and Fermi \cite{Fermi} demonstrates).  
That challenge is worsened by going to
finite-T (one must devise an entropy functional and incorporate 
the intrinsic T-dependence of other functionals). Third, the ground 
state approximations must be reliable and
well-founded if there is to be a sensible T=0 K limit for 
approximate free energy functionals. 

\subsection{Basics and Notation} 

For context and to set notation, the Levy-Lieb version of the
Hohenberg-Kohn universal functional \cite{H-K,Levy79,Lieb83} is \vspace*{-4pt}
\beq
{\mathcal E}[n] = T[n] + U_{ee}[n]
\label{HKfunc}
\vspace*{-4pt}
\eeq
with $T[n]$, $U_{ee}[n]$, and $n({\mathbf r})$, the total kinetic energy (KE), 
total Coulomb energy (Hartree, exchange, and correlation), and the electron
density at point $\mathbf r$ respectively ($\int d{\mathbf r}n({\mathbf r}) = N_e$,
with $N_e$ the number of electrons).   Assuming that the 
sum is bounded below, addition of an external 
potential energy $E_{ext}[n]$  gives the usual DFT
variational principle,  \vspace*{-4pt}
\beq
\min_n \lbrace {\mathcal E}[n] + E_{ext}[n] \rbrace = E_0[n_0] \; .
\label{HKmin}
\vspace*{-7pt}
\eeq
Zero subscripts indicate ground state values.  

The Kohn-Sham \cite{KohnSham} (KS) 
rearrangement of $\mathcal E$ invokes 
an auxiliary non-interacting Fermion system with 
the same density as the physical system.  This raises
so-called $v$-representability requirements which we assume to 
be satisfied.  
The KS system has KE and exchange (X) energies  $T_s$, $E_x$, 
which enable the regrouping of (\ref{HKfunc}) into \vspace*{-4pt} 
\bea
{\mathcal E}[n]& = & T_s[n] + E_H[n] + E_x[n] +  E_c[n]  %
\label{KSdecompfunctional}  \\
E_c[n] &:=& U_{ee}[n] - E_H[n] - E_x[n] + T[n] - T_s[n]  %
\label{Ecdefn} \\
E_H[n] &:= & \half \int d{\mathbf r}_1 d{\mathbf r}_2  %
\frac{n({\mathbf r}_1) n({\mathbf r}_2) }{|{\mathbf r}_1 -{\mathbf r}_2| } %
\label{EHdefn} \\
E_x[n] &:=& \Bra{\Phi_s[n]}\hat V_{ee}\Ket{\Phi_s[n]} -E_H[n]
\; . \label{Exdefn}
\vspace*{-4pt}
\eea
$E_c$ is the DFT correlation energy. It often is useful to 
write $E_{xc} = E_x + E_c$.  
$\hat V_{ee}$ is the electron-electron Coulomb interaction.  
The KS Slater determinant $\Phi_s[n]$ 
is comprised of the orbitals 
from the KS system Euler equations, \vspace*{-4pt}
\beq
h_{KS}[n]\varphi_i  =  \varepsilon_i \varphi_i\;\;\;,\;\;\; %
n({\mathbf r})  =  \sum_i f_i |\varphi_i({\mathbf r})|^2 \; .
\label{kseqnanddens}
\vspace*{-4pt}
\eeq
Here the $f_i = 0,1,2$ are occupation numbers in the 
non-spin-polarized case \cite{ParrYangBook,DreizlerGrossBook}.  
The KS potential is \vspace*{-4pt}
\bea
{v}_{KS} &=& {v}_H +  {v}_{xc} +  {v}_{ext}   \nonumber \\
{v}_{H}& =& \int d{\mathbf r}_2   %
\frac {n({\mathbf r}_2) }{|{\mathbf r} -{\mathbf r}_2| } %
\;\;\;,\;\;\; %
{v}_{xc} = \frac{\delta E_{xc}}{\delta n}  \;, 
\label{potdefns}
\vspace*{-4pt}
\eea
For now we leave ${v}_{ext}$ unspecified.  Finally the KS KE is 
\vspace*{-4pt}
\be
T_{s}[\{\varphi_i\}] = \half\sum_{i=1}^{N_{\rm e}} f_i \int \, d {\bf r} %
|\nabla \phi_i(\mathbf{r})|^2  %
:= \int \, d {\mathbf r}t_{orb}[n({\mathbf r})]  
\label{TsDefn}
\vspace*{-7pt}
\ee
in Hartree atomic units.  This positive-definite integrand form of $T_s$ is
preferable for OFDFT because the integrand of the ordinary 
Laplacian form of $T_s$ can have both signs.  The two forms differ by a 
surface integral which is zero for physically significant systems.

\subsection{Essential Challenge of OFDFT}

Posed succinctly, the OFDFT opportunity is that the computational 
costs of direct
minimization of Eq.\ (\ref{KSdecompfunctional}) scale 
with system size.   In contrast,  
solution of the KS equations, (\ref{kseqnanddens}), has  
computational cost scaling as $N_e^3$ or worse. 

The OFDFT challenge can be stated succinctly too.  $E_x[n]$ is {\it defined} 
in terms of the KS orbitals, hence is known exactly only as 
an implicit functional of $n$.   
 $T_s[n]$ is an implicit functional 
as well. $E_c[n]$ is {\it defined} in terms 
of those two.  One might try reversion to ${\mathcal E}[n]$ for 
construction of approximations but most rigorous knowledge 
about $\mathcal E$ (scaling, bounds, limits, etc.) is in terms of 
the KS rearrangement.  Roughly a half
century of effort has been devoted to finding good approximations
to $E_x$ and $E_{xc}$. Abandoning the KS decomposition would 
discard that resource and, worse, disconnect the result from a huge literature
of calculations with such functionals.  And $T_s$ has several 
rigorously demonstrable properties which serve as stringent constraints 
on approximations \cite{PostNAMETOFKEreview,JCAMD,Signpost,KJTH-PRB80}.

In short, retention and use of the KS decomposition is practically 
inescapable.  OFDFT thus aims at reliable approximations for KS DFT
quantities without explicit dependence on the KS orbitals.  
The allowed variables therefore are $n$ and its spatial derivatives.  
For $E_{xc}$ the consequence is a restriction to the meta-generalized-gradient
approximation (mGGA) rung of the widely quoted Perdew-Schmidt Jacobs' ladder of
complexity \cite{Perdew01}.  mGGAs depend upon $n$, 
$|\nabla n|$, $\nabla^2 n$, and the KS KE density $t_{orb}$.  Immediately 
the OFDFT challenge is in play:  an explicit functional for $t_{orb}$ is
required.  

In practice, the highest spatial derivative dependence that so far has been
useful for $t_{orb}$ is a GGA, to wit
\bea
T^{GGA}_{s}[n] &=&c_{TF}\int d{\mathbf r} \, n^{5/3} ({\mathbf r})  %
F_{t}(s({\mathbf r}))  \nonumber \\
c_{TF} & = &   \frac{3}{10}(3\pi)^{2/3} \; .
\label{TsGGA}
\eea
$F_t$ is called the enhancement factor.  For $F_t = 1$, $T_s^{GGA}=T_{TF}$, 
the Thomas-Fermi functional.  The 
dimensionless reduced density gradient is 
\beq
s:= \frac{1}{2(3\pi^2)^{1/3}} \frac{|\nabla n|}{n^{4/3}} \equiv %
\kappa \frac{|\nabla n|}{n^{4/3}}  \; .
\label{sdefn}
\eeq
Remark: The $s$ variable occurs in GGA X functionals also.  They
have the same form as Eq.\ (\ref{TsGGA}) but with $n^{4/3}$ rather
than $n^{5/3}$, $F_x$ rather than $F_t$, and a different 
prefactor, $c_x = -(3/4)(3/\pi)^{1/3}$.   

Eq.\ (\ref{TsGGA}) is a one-point GGA functional.  In the OFKE literature 
there is extensive work on two-point functionals, generically
$\int d{\mathbf r} d{\mathbf r}^\prime n^\alpha({\mathbf r}) %
K({\mathbf r},{\mathbf r}^\prime) n^\beta({\mathbf r}^\prime) $.  
See Section 2.3 of Ref.\ \cite{PostNAMETOFKEreview} for brief discussion and references.
One readily can imagine constructing a GGA for two-point functionals
but we are unaware of effort along that line. Instead the emphasis has been
on constructing $K({\mathbf r},{\mathbf r}^\prime)$ via constraints, 
mostly to match response properties of the weakly perturbed homogeneous 
electron gas.

Motivated to optimize computational performance, our group has focused 
on finding and exhausting the limits of single-point GGAs for $T_s$.   
The remaining discussion assesses what we have found, with a focus
on the surprising non-universality of approximate OFKE functionals, 
implications for their common use with external
potentials of regularized Coulomb form, and an empirical attempt to 
ascertain the limits of GGA performance for a particular kind of
regularized potential. 

\section{Qualitative Distinctions among GGAs for $t_{orb}$}

Our work began \cite{JCAMD} by testing multiple published $T_s^{GGA}$  
functionals.  A rough classification introduced then was 
standard GGA and modified-conjoint GGA (mcGGA).  The latter term stems
from conjoint functionals \cite{LeeLeeParr91}, i.e.\ those for which 
$F_t \propto F_x$.  Standard GGAs include
the second-order gradient approximation (SGA)
\bea
T^{SGA} &=& T_{TF} + \frac{1}{9}T_W \label{TSGA} \\
T_{W}[n] &:= &%
\frac{1}{8}\int d{\mathbf r}  \frac{|\nabla n({\mathbf r})|^2} {n({\bf r})} %
\label{TWdefn}
\eea
and the von Weizs\"acker KE $T_W$ itself, along with most of the
GGAs of the modern era, e.g.\ that by Perdew \cite{Perdew92}, the 
PW91 KE functional based on the Perdew-Wang X functional \cite{LacksGordon93}, 
and those from  DePristo and Kress \cite{DePristoKress87}, 
Thakkar \cite{Thakkar92}, and Tran and Wesolowski \cite{TranWesolowski02}, and 
the APBEK functional based on the PBE X functional \cite{CFDS11}. 
Modified conjoint GGAs arise from altering or refining the conjointness
conjecture (which is not strictly correct; see Ref.\ \cite{KJTH-PRB80}).
These include $T_{TF}+ T_W$ and our PBE2, KST2 \cite{JCAMD,KJTH-PRB80},
and VT84F \cite{PRB88-161108R} functionals.

The two functional types have qualitatively different
performance. Ordinary GGAs predict the KE order of magnitude correctly
but fail to give binding for simple molecules and solids.  There
are some exceptions for solids in which a pseudo-density is used.  
mcGGAs do bind simple molecules 
as well as many solids at least semi-quantitatively, 
but they overestimate the KE strongly. 
As a consequence the total energy also is strongly overestimated (too 
high).  
We also have found that these functionals exhibit peculiar 
sensitivity to the type of pseudo-potential used, behavior 
found by others as well \cite{1408-4701}.  

The main difference between ordinary GGAs and the mcGGAs is the
enforcement of positivity constraints on the mcGGAs.  Enforcement
is via imposition of requirements upon the density in the
case that the external potential is Coulombic, 
\be
{v}_{ext}({\mathbf r})  =  -\sum_\alpha \frac{Z_\alpha} %
{|{\mathbf r}-{\mathbf R}_\alpha|} 
\label{vext-coulomb}
\ee
with $Z_{\alpha}$ the atomic number of the nucleus at site
${\mathbf R}_\alpha$.   Such approximate 
$T_s^{GGA}$ functionals therefore are not guaranteed to be universal, 
even though $T_s$ is. Two questions then arise.  Are both the overly
large KE and sensitivity to pseudo-potentials of mcGGAs  connected
with this non-universality?  Is there an example of a $T_s^{GGA}$ 
that has both the good KE magnitudes of an ordinary GGA and the 
good binding properties of an mcGGA?  We address these two 
issues in the remainder of this paper.

\section{Positivity and Near-origin Conditions}

The Pauli-term decomposition
\beq
T_{s}[n] = T_{W}[n]+ T_{\theta}[n]  \;\;\;,\;\;\;  %
\label{PauliDecomp}
\eeq
provides a rigorous bound \cite{TalBader78,BartolottiAcharya82,%
Harriman87,LevyOu-Yang88,Baltin87},
\beq
T_{\theta}[n] \; \geq 0 \; ,
\label{Tthetapositive}
\eeq
because $T_W$ is a lower bound to the KS KE 
\cite{TwoHoffman-Ostenhofs77,SearsParrDinur80,Harriman85,Herring86}.
$T_W$ also is the exact $T_s$  for one electron, a fact that will become useful
shortly. (It also is exact for a two-electron singlet.)
The Pauli term potential also is rigorously non-negative: 
\beq 
{v}_\theta({\mathbf r}) := \frac{\delta T_\theta[n]}{\delta n({\mathbf r})} \geq 0 \;%
\;\;,\;\; \forall {\mathbf r}  \;\;.
\label{vthetapos}
\eeq

These are universal properties of $T_s$. For a GGA, the Pauli
term separation corresponds to a  $T^{GGA}_\theta$ with energy
density $t_\theta$ and enhancement factor 
\begin{equation}
F_{\theta}(s) = F_{t}(s) - \frac{5}{3}s^2 \; .
\label{Ft2}
\end{equation}
Though $T_\theta \ge 0$, it is not necessarily the case that the 
associated Pauli-term energy density $t_\theta$ obeys the same
positivity $t_\theta \ge 0$ because energy densities are defined
only up to additive functions which integrate to zero.  Refs.\ 
\cite{LevyOu-Yang88,Herring86,LevyPerdewSahni84} chose the
canonical form for $t_\theta$ (i.e. that which comes from the KS
equation), which is positive semi-definite.  We 
adopted that argument in Ref.\ \cite{TKJ-IJQC109}.  The
consequence, to which we return in Sect.\ 4, is
\beq
F_{\theta}(s({\mathbf r})) \geq 0 
\;,\;\; \forall {\mathbf r}  \;\;.
\label{FthetaPos}
\eeq

To have enough additional constraints to determine a useful approximate 
$F_\theta$, 
we used \cite{PRB88-161108R,JCAMD,KJTH-PRB80,TKJ-IJQC109} 
requisites of physical many-electron systems, i.e.,  
those with an external potential given by Eq.\ (\ref{vext-coulomb}).
Non-universality enters.  

The nuclear-cusp condition \cite{Kato57} density
\beq
n(r)\sim e^{-2Zr}=(1-2Zr)+O(r^2)\,.
\label{E4}
\eeq
gives ${v}_{\theta}^{GGA}(r)\sim a/r$  where $a$ is a
constant which depends on the specific enhancement factor \cite{KJTH-PRB80}.
So far as we know, the first mention of this consequence 
was by Levy and Ou-Yang; see the latter part
of Section III of Ref.\ \cite{LevyOu-Yang88}

The one-electron character of the tail region of a many-electron atom
\cite{LevyPerdewSahni84} forces the approximate functional to 
go over to $T_W$ in that region \cite{DreizlerGrossBook}.  For a GGA therefore, 
\beq
\lim_{s\rightarrow \infty} F_{\theta}(s)= 0 \; .
\label{largeslimFt}
\eeq

Kato cusp behavior Eq.\ (\ref{E4}) is not exhibited by   
any density that results from a regularized potential, e.g., a
pseudo-potential.  See for example, Eq.\ (6) in Ref.\ 
\cite{LambertClerouinZerah06} and associated discussion.
Removing that cusp to allow use of compact basis sets (especially
a plane-wave basis) is the motive for pseudo-potentials.  
Densities from expansion in a finite Gaussian-type basis set,
even in all-electron calculations that use Eq.\ (\ref{vext-coulomb}), also
do not have Kato cusp behavior.   
Similarly, the proper tail behavior, also exponential, is not found in any
finite Gaussian expansion density. 
Here we focus on the former issue, the near-origin behavior of atomic-like 
systems.

Consider one-center $N_e$-electron densities of the flexible form
\bea
n_f({\mathbf r}) &:=& A_f \exp (-\lambda r^\gamma) \; , \; 1\le\gamma\le2 
\label{nfdefn} \\
A_f &=& \frac{N_e \gamma \lambda^{3/\gamma}}{4\pi\Gamma(3/\gamma)} \; ,%
\label{nfnorm} 
\eea
The norm follows from Ref.\ \cite{GS3478}, with
the usual $\Gamma$ function.  
With $\gamma =1$, $\lambda = 2N_e$, $N_e=1$, 
this is the H atom density in the central field approximation.  
For $\gamma =2$ it is pure Gaussian. For use in what follows, 
the von Weizs{\"a}cker potential for densities of this form is 
\be
v_W = \frac{\delta T_W}{\delta n} = \frac{\lambda\gamma}{8}r^{\gamma -2} %
\lbrack 2 (\gamma + 1) -\lambda\gamma r^\gamma \rbrack  \;  .
\label{vWpotfornsubf}
\ee

With densities of the form (\ref{nfdefn})
we can explore two simple but illuminating issues.  
The first is to determine the external potential 
that corresponds to the given density
for the case $N_e=1$. Recall the bijectivity of the external
potential and the density guaranteed by the first Hohenberg-Kohn 
theorem \cite{H-K}.   The central-field 
hydrogenic case is obvious but it is
instructive to do it in the context of OFKE functionals. The Euler equation is
\be
\frac{\delta ({\mathcal E} + E_{ext})}{\delta n}  =   v_W + v_\theta %
+ v_H + v_{xc} + v_{ext} = \mu  \; ,
\label{OFgenericEuler}
\ee
with $\mu$ the Lagrangian multiplier for charge normalization. 
$T_W$ is exact for the one-electron case, so $v_\theta = 0$.  Exact exchange
cancels the Hartree self-interaction, so $v_H = - v_x$, and there is no 
correlation, $v_c =0$.
The von Weizs\"acker potential (\ref{vWpotfornsubf}) for the 
hydrogenic densities ($\gamma=1$, $\lambda = 2N_e$) is 
\be
v_W = \frac{N_e}{r} - \frac{N_e^2}{2} \; .
\label{vWpot}  
\ee
For H, $\mu = -\half$, $N_e=1$, (\ref{OFgenericEuler}) gives the
expected result:
\be
0  =  \frac{1}{r} - \frac{1}{2} + v_{ext}(r) - (-\half) \;\;\; %
\Rightarrow \;\;\; v_{ext}(r) = -\frac{1}{r}   \; . 
\label{Hvext}
\ee

Redoing the argument with $\gamma =2$, $N_e=1$ gives
\be
v_W = \frac{\lambda}{2} ( 3 - \lambda r^2) %
\;\;\; \Rightarrow \;\;\; %
v_ {ext} = \half \lambda^2 r^2 + (\mu - \frac{3\lambda}{2}) \; , 
\label{vextlambdaeq2}
\ee
the expected quadratic dependence for $v_{ext}$.

This elementary exercise illustrates a significant point for 
approximate functionals. Repeat the argument for $\gamma =2$ but
now with the physically important external Coulomb potential imposed 
and with an approximate $T_\theta$ functional (not necessarily a GGA;
for the moment the discussion is general).
Then the Euler equation becomes 
\be
\mu = -\frac{1}{r} - \half \lambda^2 r^2 + \frac{3\lambda}{2} + 
v_\theta^{approx} (r) \; .
\label{Eulermismatch}
\ee
The only way this can be satisfied is for there to be an {\it incorrect},
i.e. non-zero, $v_\theta^{approx}$ for the one-electron case.  

In the case of pseudo-potentials, the argument runs in reverse.  
Suppose a pseudo-potential prescription to be used at the so-called 
one-electron level, i.e., one electron outside the core, and suppose
it to deliver the form (\ref{vextlambdaeq2}). Assume that one can 
contrive a satisfying approximate functional with the property that for 
$N_e=1$, the approximate functional respects rigorous constraints for the 
corresponding pseudo-density.  Now shift to an all-electron pseduo-potential
and shrink the core toward the bare Coulomb potential.  In an arbitrarily
small region around the origin, the pseudo-density
will remain harmonic but the pseudo-potential in almost all space will be
essentially Coulombic, leading to the kind of  mismatch given in 
Eq.\ (\ref{Eulermismatch}).  Even at this level (two pseudo-potentials
with the same regularization procedure but significantly different
core radii and populations), there is a 
lack of universality for the approximate $T_\theta$.

For arbitrary $\gamma$ dependence,  $1 \le \gamma \le 2$, the imputed 
external potential is
\be
v_{ext} = \mu - v_W = \mu + \frac{\lambda\gamma}{8}r^{\gamma -2} %
\lbrack \lambda\gamma r^\gamma - 2 (\gamma + 1) \rbrack  \;  ,
\label{vextflexgeneric}
\ee
(with suitably adjusted $\mu$ of course). 
Similar mismatch difficulties will occur for all intermediate
$\gamma$ values, as will the singularities for $\gamma \ne 2$. 

Now consider GGA functionals with arbitrary $N_e$.  The GGA  
Pauli potential is \cite{KJTH-PRB80}
\be
v_{\theta}^{\rm GGA}(s^2) =c_0 n^{2/3} %
\left\lbrace \frac{5}{3}F_\theta (s^2) %
- \left(\frac{2}{3}s^2 + 2p \right) \frac{\partial F_\theta}{\partial (s^2)} %
\right. 
 \left. %
+ 4\left(\frac{4}{3} s^4 - q \right) \frac{\partial^2 F_\theta}{\partial %
 (s^2)^2} \right\rbrace \, ,
\label{vthetas2pq}
\ee
with higher-order reduced density derivatives 
\be
p :=\kappa^2 \frac {\nabla^2 n}{n^{5/3}} \;\;\; , \;\;\; %
 q := %
\kappa^4 \frac{\nabla n \cdot (\nabla \nabla n) \cdot \nabla n}{n^{13/3}} %
\; .
\label{pqdefn}
\ee
Evaluation with the flexible density (\ref{nfdefn}) yields 
\be
s^2(r) = \kappa^2 \lambda^2 \gamma^2 r^{2(\gamma-1)}n_f^{-2/3}(r)  \; , 
\label{ssquarednf}
\ee
\be
p(r) = \kappa^2\lambda\gamma r^{\gamma-2} %
\lbrack \lambda \gamma r^\gamma - (\gamma + 1) \rbrack n_f^{-2/3}(r)  \; , 
\label{pnf}
\ee
and 
\be
q(r) = \kappa^4 \lambda^3 \gamma^3 r^{3\gamma-4} %
\lbrack \lambda \gamma r^\gamma - (\gamma -1) \rbrack  n_f^{-4/3}(r)  \; . 
\label{qnf}
\ee
%

Except for a negative sign, the coefficient 
of $\partial F_\theta/\partial s^2$ in Eq.\ (\ref{vthetas2pq}) is 
\be
\frac{2}{3} s^2 +2p = %
2 \kappa^2 \lambda \gamma r^{\gamma -2} 
\lbrack {4\over 3} \lambda \gamma r^\gamma - (\gamma +1) \rbrack n_f^{-2/3}(r)  \; . 
\label{coeffdfds2A}
\ee
Notice the singularity at the origin for $\gamma < 2$.  
Up to a factor of $4$, the coefficient of $\partial^2 F_\theta/\partial (s^2)^2$ in
Eq.\ (\ref{vthetas2pq}) is 
\be
\left\lbrack \frac{4}{3} s^4 - q \right\rbrack = %
\kappa^4 \lambda^3 \gamma^3 r^{3\gamma-4} %
\left\lbrack \frac{\lambda\gamma}{3} r^\gamma + (\gamma -1) \right\rbrack %
n_f^{-4/3} \; .
\label{coeffd3fds22A}
\ee
This is non-singular only for $\gamma \ge 4/3$.  

For small $s$, one usually enforces gradient expansion behavior on
$F_\theta$, 
\be
F_\theta = 1 + as^2  \; 
\label{fthetaGE}
\ee
and only the first derivative term in $v_\theta$, Eq.\ (\ref{vthetas2pq}), 
is at issue.  After a bit of manipulation, 
\be
p(r) = s^2 \left\lbrack 1 - %
\frac{\gamma + 1}{\lambda \gamma r^\gamma} \right\rbrack \;\;\; \Rightarrow %
\;\;\; \frac{2}{3} s^2 +2p = 2s^2 \left\lbrack \frac{4}{3} -  %
\frac{\gamma + 1}{\lambda \gamma r^\gamma} \right\rbrack  \; .
\label{coeffdfds2B}
\ee
The singularity structure in $v_\theta$ then is evident.  The
general result is 
\be
v_\theta^{GGA}[n_f]=c_0n_f^{2/3}\left\lbrace \frac{5}{3} + %
a\kappa^2\gamma^2\lambda^2 r^{2(\gamma-1)}n_f^{-2/3} %
\left\lbrack \frac{2(\gamma + 1)}{\lambda \gamma r^\gamma} -1 %
\right\rbrack \right\rbrace \; .
\label{vthetaGGAnfgen}
\ee
For convenience the two limiting cases are 
\be
v_\theta^{GGA}[n_f,\gamma=1]=c_0n_f^{2/3}\left\lbrace \frac{5}{3} + %
a\kappa^2\lambda^2 n_f^{-2/3} \left\lbrack \frac{4}{\lambda r} -1 %
\right\rbrack \right\rbrace 
\label{vthetaGGAgammaeq1}
\ee
and
\be
v_\theta^{GGA}[n_f,\gamma=2]=c_0n_f^{2/3}\left\lbrace \frac{5}{3} + %
4a\kappa^2\lambda n_f^{-2/3} \left\lbrack 3 - \lambda r^2 %
\right\rbrack \right\rbrace \; .
\label{vthetaGGAgammaeq2}
\ee
The takeaway point is that if one sets out to build an approximation 
constrained to behave properly for $\gamma =1$ (the physical case), 
the singularity is inevitable and the near-origin positivity is 
determined by the sign of the gradient expansion coefficient $a$.
Our mcGGAs are built to have $a > 0$.  
However, if the actual density is regularized and has Gaussian form
near the origin, then if that density is ``cuspy'' enough, 
i.e. has large $ \lambda $, even with $ a > 0 $ and positivity 
constraints enforced on building the approximation, there still 
can be small-$r$ regions for which $v_\theta^{approx} < 0 $.  

\section{Empirical Exploration of the Limits of GGA KE}
\subsection{Methodology}

Our approach to the development of  GGA OFKE functionals 
has been to adopt some suitable analytical form for the kinetic energy 
enhancement factor $F_t$ with a few parameters determined
from imposing constraints
(e.g.\ correct scaling if applicable, 
correct small-$s$ and large-$s$ behavior)  
and, if unavoidable, fitting to a small set of training data. 
Interpolation 
between small- and large-$s$ is defined by the chosen analytical form
for $F_t$. The analytical forms usually are relatively simple with
deliberately limited flexibility to avoid 
introduction of non-physical kinks or other artifacts in that
interpolation.  In this
sense, the properties are analogous to those   
of standard finite basis sets (see for example 
Refs.\ \cite{Perdew92,DePristoKress87,Thakkar92,TranWesolowski02,JCAMD,Signpost,PRB88-161108R}).

The unwelcome effects of limited flexibility can be avoided, at least
in principle, by use of a numerical enhancement factor given  
on a mesh $s_0=0$, $s_1$, ... , $s_n=s_{max}$ 
There is a practical barrier however. To determine such a numerical
$F_t$ requires numerical integration in real space of the complicated
$(n,|\nabla n|)$ dependence of the kinetic energy functional integrand 
$t_{orb}$, Eq.\ (\ref{TsDefn}), evaluated on a numerical $s$-mesh. 
Experience demonstrates that the 
result is unphysical, noisy, numerically unstable results. One can
see the difficulty simply by considering the numerical integration of
an $s$ ``density of states'' on a mesh of points $s_i$:
\be
{\mathcal D}(s_i) := \int d{\mathbf r} \delta(s_i - s({\mathbf r})) \approx %
\sum_j w_j \delta_{s_i,s(\mathbf{r}_j)}  \; ,
\label{Dofs}
\ee
with $w_j$ the quadrature weights. Numerical experiment shows that 
a modest change in even a very fine $\mathbf r$ or $s$ mesh (or both)
leads to distinctly different results.     

An effective alternative is Pad\'e approximants \cite{Pade} of high orders 
such as were 
used recently for analytical representation   
of common Fermi-Dirac integral combinations \cite{FD.2015}.
They provide the simultaneous flexibility and smoothness
required by the numerical integration in $\mathbf r$.
Numerical exploration led to the Pad\'e approximant 
\be
F_{{t}}(s) = \frac{1+\sum_{i=1}^{k}a_is^i}
{1+\sum_{i=1}^{l}b_is^i}
\,,
\label{Ft}
\ee
of order [9,10] in the variable $s$ ($k=9$, $l=10$)
as a workable compromise between flexibility and 
number of free parameters.

Only a few parameters in the approximant can be determined 
from imposition of constraints. The remainder must be obtained
by fitting.  For the present study, the {\it only} constraint  
imposed on Eq. (\ref{Ft})  is recovery of the correct 
second-order gradient expansion at small $s$, 
\be
F_{{t}}(s)\approx 1+\frac{5}{27} s^2, ~~~ s<<1 
\,.
\label{Ft-small-s}
\ee
This is accomplished by setting $a_1=b_1$, $a_2=(5/27)+b_2$.
To allow maximal freedom for the fitted $F_t$ we have not imposed
the large-$s$ von Weizs\"acker limit given in Eq.\ (\ref{largeslimFt}).
Studies of X GGA functionals \cite{PBELS} show that the distribution of $s$,
Eq.\ (\ref{Dofs}), is negligible above about $s=3$ for most systems of
interest, thus suggesting that such large-$s$ behavior
constraints are not critically important, at least for fitting.  
Also note well that in what follows, we have {\it not} imposed 
{\it any} of the
positivity constraints, Eqs.\ (\ref{Tthetapositive}), (\ref{vthetapos}), and (\ref{FthetaPos}).  The motivation is to make the empirical fitting as
unconstrained as possible.

The absolute KE versus binding energy dilemma posed at 
the end of Sect.\ 2 has implications for the fitting criteria to be used.
The usual {\it KE} fitting criterion is equivalent to the
{\it total energy} or $E$ criterion  \cite{JCAMD}, namely to minimize 
the squared energy difference 
between non-self-consistent OFDFT and reference KS energies at system 
equilibrium geometries (from standard KS calculations).  
The procedure is non-self-consistent  for the OFDFT calculations
because KS densities are used as input.
The obvious
flaw in the $E$ criterion (which was investigated in Ref.\ \cite{KJTH-PRB80}) 
is that it forces OFDFT total
energies at KS equilibrium configurations to be as nearly correct as possible
but ignores the shape of the KS binding energy curve near equilibrium.
    
In Refs.\ \cite{JCAMD,Signpost,KJTH-PRB80} Frank Harris and two 
of us introduced what was called the $\Delta E$ criterion.  In it, the 
objective function to be optimized is formed from energy differences 
between a point away from equilibrium and the equilibrium point as 
predicted by the reference conventional KS calculation.  The objective function
has two of those energy differences, one from OFDFT, the other from the KS
calculations. Obviously the $\Delta E$ criterion enforces binding upon the 
OFDFT approximation while leaving the total energy uncontrolled.  The  
result can be an excessively high total energy.  

In the present work we address 
these two limitations by making a convex sum of average 
versions of the two criteria. The averages are calculated 
over all atoms, molecules (and 
their geometries) in a training set.
To put the two criteria on the same scale, we use the 
mean absolute relative error (MARE) of energy differences rather than 
average absolute energy differences,     
\be
\omega_{\Delta E}={1\over N}\sum_{M,i\ne e} 
\frac{
\left|\,\Delta E^{\rm KS}_{M,i}-
\Delta E^{\rm OF\mbox{-}DFT}_{M,i}\right|}
{\left|\Delta E^{\rm KS}_{M,i}\right|}
\,.
\label{omega-DeltaE}
\ee
Here, for the nuclear spatial configuration $i$ of  molecule $M$, 
$\Delta E_{M,i}= E_{M,i}-E_{M,e}$, with $E_{M,e}$ the energy associated 
with the equilibrium nuclear configuration as predicted from 
conventional KS computations, and $N$ the total number of terms in the 
sum Eq.\ (\ref{omega-DeltaE}).
Similarly, the mean absolute relative error 
of the total energy is 
\be
\omega_{E}={1\over N}\sum_{M,i} 
\frac{
\left|\,E^{\rm KS}_{M,i}-E^{\rm OF\mbox{-}DFT}_{M,i}\right|}
{\left|E^{\rm KS}_{M,i}\right|}
\,.  
\label{omega-E}
\ee
The objective function is a convex combination of both
\be
\omega(\alpha)=\alpha \omega_{E} + (1-\alpha) \omega_{\Delta E}
\,,
\label{omega}
\ee
with $\alpha\in [0,1]$. Minimization of $\omega(0)$ is essentially the 
$\Delta E$ criterion, and conversely for $\omega(1)$.  One expects,
or at least hopes,  that some intermediate $\alpha$ will provide 
a KE functional with both reasonable binding and reasonable 
absolute energy errors.

The training set we used includes nine molecules comprised of first-
and second-row atoms and of diverse bonding types along with three
closed shell atoms, 
$M=\{{\rm LiH, CO, N_2,}$ ${\rm LiF, BF, NaF, SiO, H_4SiO, H_4SiO_4, Be, Ne, Ar}\}$.  A set of six bond lengths was
used for each molecule. Molecular geometries were changed by varying
the single bond length in the diatomics, the central bond length
$R$(Si--O) in H$_4$SiO, and by varying $R$(Si--O$_i$) in H$_4$SiO$_4$
deformed in the $T_d$ mode.  This set is small by comparison with the
training sets used in the Minnesota series of XC functionals
\cite{Minnesota} because our purpose is different.  We do not seek a
broadly useful empirical functional.  The issue here is narrower,
namely whether there exists an OFKE functional which does well both on
absolute energies and binding even on a small sample of systems.

One other technical point is that the enhancement factor 
$F_t$ which results from fitting is checked for poles on the interval 
$s\in [0,1000]$.  If the denominator of Eq.\ (\ref{Ft}) has a root 
on that interval, the corresponding set of parameters is rejected.

All reference KS calculations were done in the local density 
approximation (LDA) for XC (see Refs.\ 
\cite{Slater2a,Slater2b,Slater2c,Gaspar,KohnSham65a,KohnSham65b,VWN80,CeperleyAlder80})
using a triple-zeta Gaussian-type basis with
polarization functions (TZVP) \cite{Ahlrichs92,Ahlrichs94,Basis}.
Orbital-free kinetic energy integrals
were calculated by numerical quadrature, as in our previous work \cite{JCAMD}.
Weight functions, $\mathrm{w}_I(\mathbf{r})$, localized near each center with the 
properties that $\mathrm{w}_I(\mathbf{r})\ge 0$ and 
$\sum_I\mathrm{w}_I(\mathbf{r})=1$ are used to represent the multicenter
integrals exactly as a sum of atom-centered contributions  \cite{Becke88}
\be
T^{GGA}_{s}[n] =\sum_{I=1}^{N_{\mathrm{atoms}}} c_{TF}\int d{\mathbf r} \, 
\mathrm{w}_I(\mathbf{r}) n^{5/3} ({\mathbf r})  %
F_{t}(s({\mathbf r}))
\,.
\label{fuzzy}
\ee
Radial integration of the resulting single-center forms was via 
a Gauss-Legendre procedure, while integration over the angular variables
used high-order quadrature formulae \cite{Lebedev99}. A dense mesh 
consisting of 150 radial and 434 angular 
grid points was used to calculate atom-centered integrals.
These computations used routines developed by
Salvador and Mayer  \cite{SalvadorMayer04} and included
in their code {\sc fuzzy}.

Before proceeding to results, one should note the implications of
these numerical procedures.  The finite Gaussian-type basis inexorably
yields Gaussian near-origin behavior of the density.  Yet the calculations
are all-electron in the bare Coulomb external potential, (\ref{vext-coulomb}).
This is precisely the inconsistency between external potential and
near-origin density behavior discussed in Sect.\ 3. 

\subsection{Results}

\begin{figure}
\includegraphics[width=8.0cm]{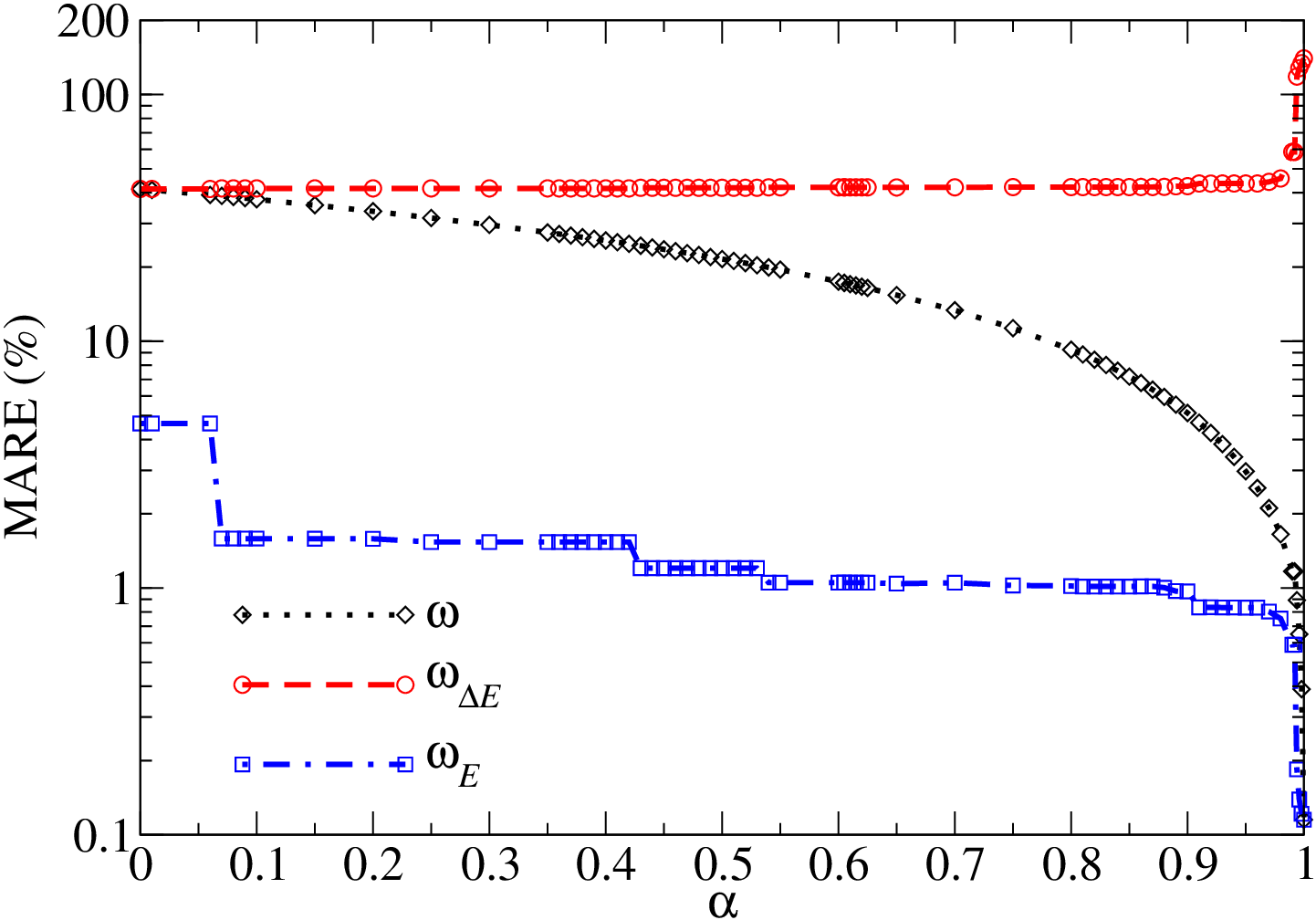}
\caption{
Minimum $\omega(\alpha)$ and 
corresponding $\omega_{\Delta E}$ and $\omega_{E}$ values as functions of 
$\alpha$.
}
\label{fb0-fb1}
\end{figure}

There are seventeen independent parameters left in Eq.\ (\ref{Ft})
after constraining to the second-order gradient expansion.  Those 
were optimized to 
minimize the objective function $\omega(\alpha)$. 
Figure \ref{fb0-fb1} shows the $\omega_{E}$ and $\omega_{\Delta E}$
MAREs, Eqs.\ (\ref{omega-E}) and (\ref{omega-DeltaE}) respectively, 
as functions of $\alpha$. The minimum $\omega(\alpha)$ value
also is shown.  It decreases monotonically from 41\% to 0.12\%.
Up to about $\alpha = 0.97$, $\omega_{E}$ decreases slowly with a few jumps
(from about  5\% to 0.8\%),
while $\omega_{\Delta E}$ is almost flat (from 41\% to 44\%). 
Unsurprisingly, they diverge as $\alpha \rightarrow 1$ 
($\omega_{\Delta E}=140\%$, $\omega=0.12\%$),
an illustration of the absolute energy vs.\ binding energy dilemma. 

\begin{figure}
\includegraphics*[height=5.4cm]{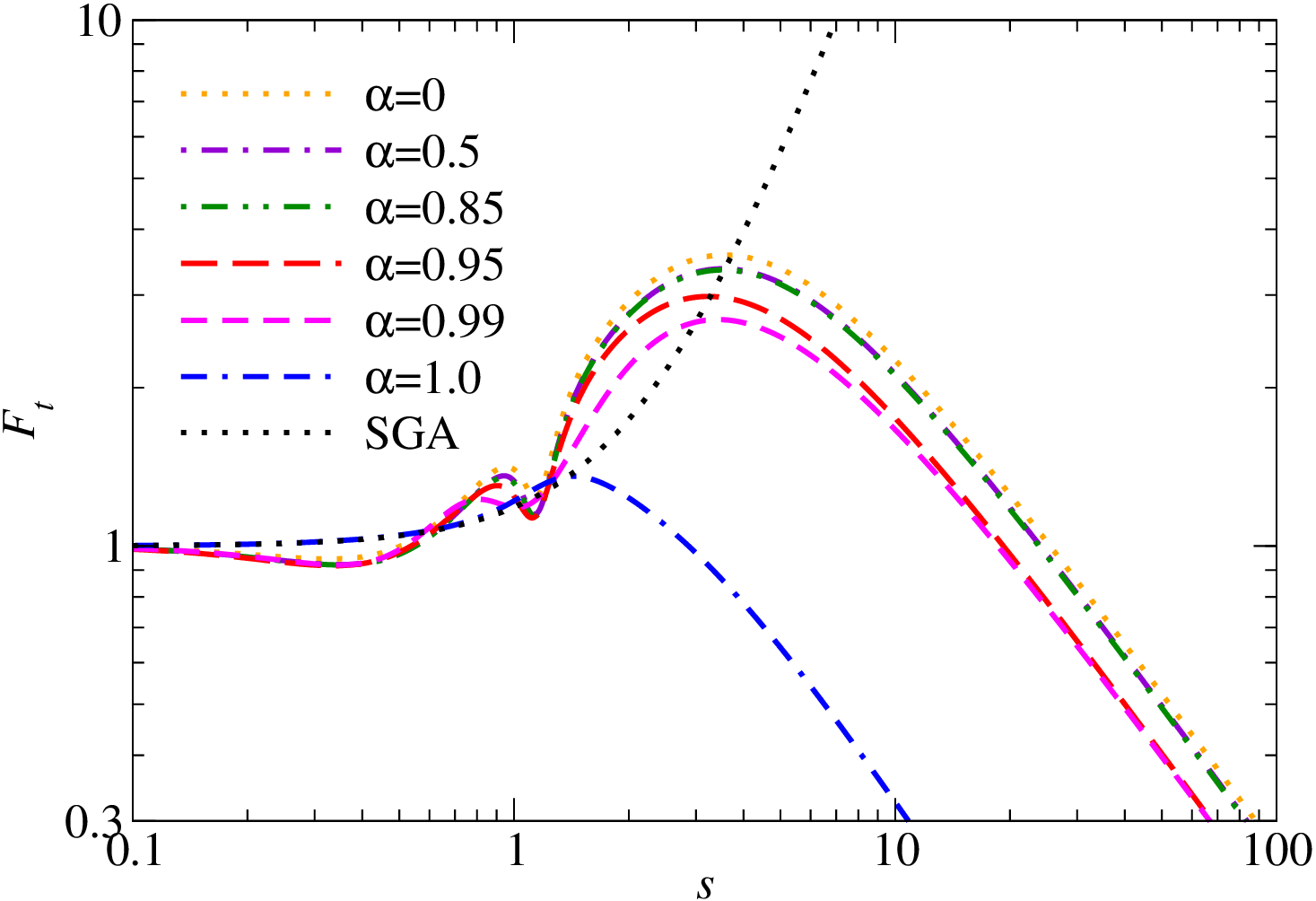}
\includegraphics*[height=5.4cm]{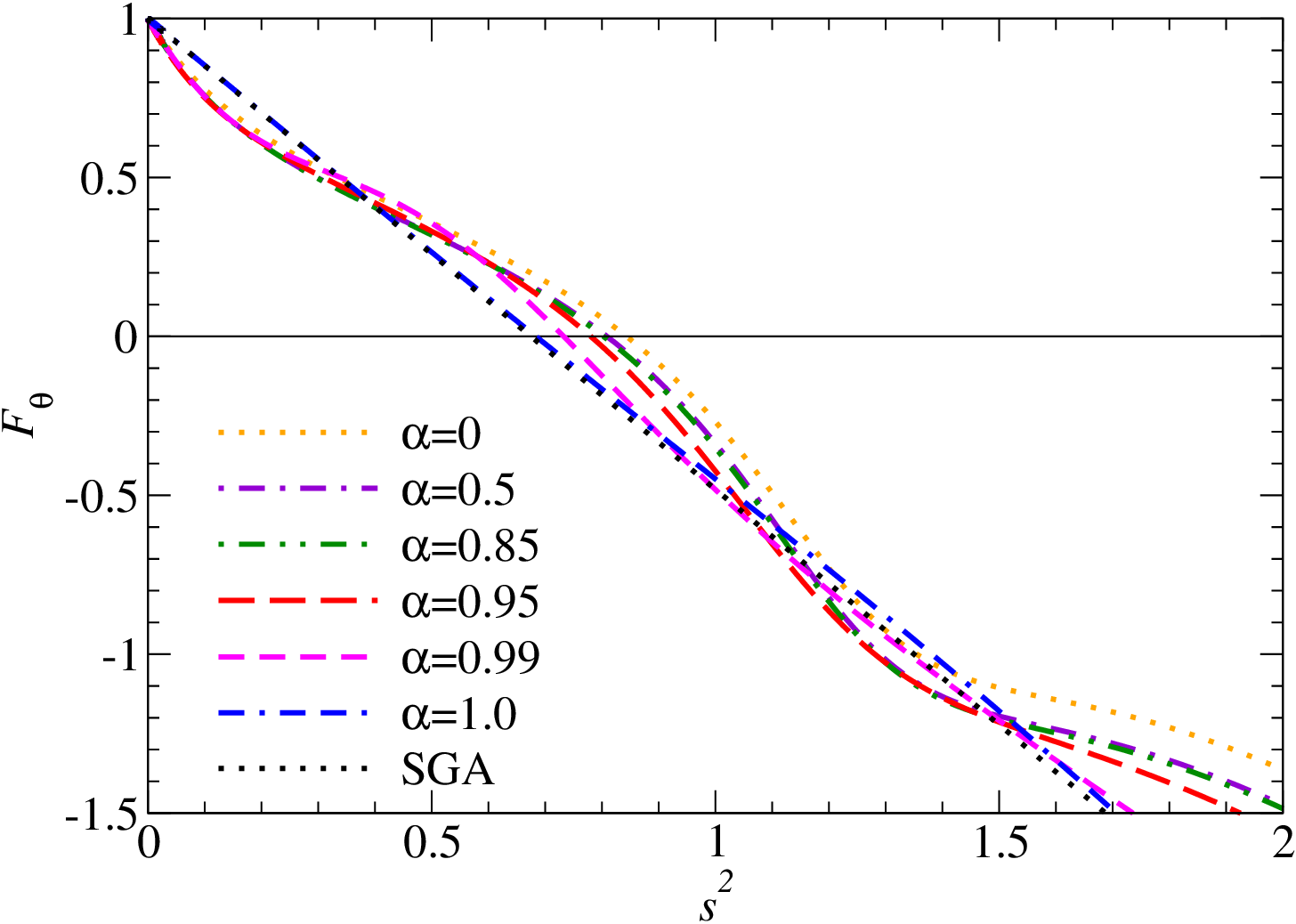}
\caption{
Non-interacting kinetic energy (left) and Pauli term (right)  
enhancement factors, $F_{t}$ and $F_{\theta}$,  
as functions of $s$ and $s^2$ respectively.
}
\label{Ft-Ftheta}
\end{figure}

Figure \ref{Ft-Ftheta} shows the fitted $F_t$ and $F_{\theta}$
for selected $\alpha$ values. Notice the violation of $F_\theta \ge 0$,
a consequence of the unconstrained fitting.  Notice also the structure
in $F_t$, especially for large $\alpha$ values.  

The right-hand panel of Fig.\ \ref{Ft-Ftheta} clearly shows the separation of 
enhancement factors into two groups corresponding to $\alpha\le0.99$ 
and $\alpha=1.0$.  The $\alpha=1.0$ curve is almost a 
line ($F_{\theta}$ is shown as a function  of $s^2$) and is practically 
indistinguishable from the SGA curve (shown for comparison) for $s<1.5$.
$F_{\theta}$ (and $F_{t}$) 
has some structure (oscillations around the SGA curve)
for $s\le 1.5$ and $\alpha\le 0.99$, 
though that may change upon changing the training set and/or 
the analytical form for the enhancement factor. 
Also, during the optimization process we noted 
the existence of many significantly different enhancement factors 
which cannot be discriminated clearly by the objective 
function $\omega(\alpha)$.

\begin{table}[htb]
\caption{\label{tab:table1}
MARE values $\omega_{E}$ and $\omega_{\Delta E}$ (in \%)
calculated individually for each system, with 
parameters from minimization of $\omega(\alpha=0.95)$
and $\omega(\alpha=1.0)$.}
\label{table:1}
\newcommand{\m}{\hphantom{$-$}}
\newcommand{\cc}[1]{\multicolumn{1}{c}{#1}}
\renewcommand{\tabcolsep}{2pc} 
\renewcommand{\arraystretch}{1.2} 
\begin{tabular}{@{}lccccc}
\hline
&\multicolumn{2}{c} {$\alpha=0.95$}&& \multicolumn{2}{c} {$\alpha=1.0$}\\
\cline{2-3}\cline {5-6}
System & $\omega_{E}$ & $\omega_{\Delta E}$ && $\omega_{E}$ & $\omega_{\Delta E}$\\
\hline
LiH          & 4.2  & 85 && 0.2 & 85\\  
CO           & 0.2  & 28 && 0.3 & 150\\ 
N$_2$        & 0.2  & 35 && 0.3 & 150\\ 
LiF          & 0.1  & 80 && 0.1 & 200 \\
BF           & 0.1  & 15 && 0.1 & 160\\ 
NaF          & 0.1  & 25 && 0.03 & 180\\ 
SiO          & 0.02 & 32 && 0.04 & 120 \\
H$_4$SiO$_4$ & 0.3  & 41 && 0.04 & 120\\ 
H$_4$SiO     & 0.02 & 50 && 0.6 & 110 \\ 
Be           & 4.0  & -- && 0.2 & --\\   
Ne           & 0.1  & -- && 0.02 & --\\  
Ar           & 0.7  & -- && 0.1 & --\\   
\hline
average         & 0.83  & 44 && 0.12 & 140 \\
\hline
\end{tabular}\\[2pt]
\end{table}

Any choice of $0.1 \le \alpha\le 0.98$ 
corresponds to the required KE functional, namely one which provides  
semi-quantitative binding, $\omega_{\Delta E}\approx 44$\%, and 
reasonable absolute energy error, ($\omega_{E}\approx 1$\%).  
Table \ref{tab:table1} lists the $\omega_{E}$ and $\omega_{\Delta E}$
values for the systems from the training set corresponding to the optimized 
$\omega(\alpha=0.95)$. The highest $\omega_{\Delta E}=85$\%
corresponds to LiH. Values for the optimized $\omega(\alpha=1.0)$
are also shown for comparison. All $\omega_{\Delta E}$ values
for $\alpha=1.0$ (except the LiH molecule)
are between 110\% and 200\%, signifying no binding.

\begin{figure}
\includegraphics*[width=8.0cm]{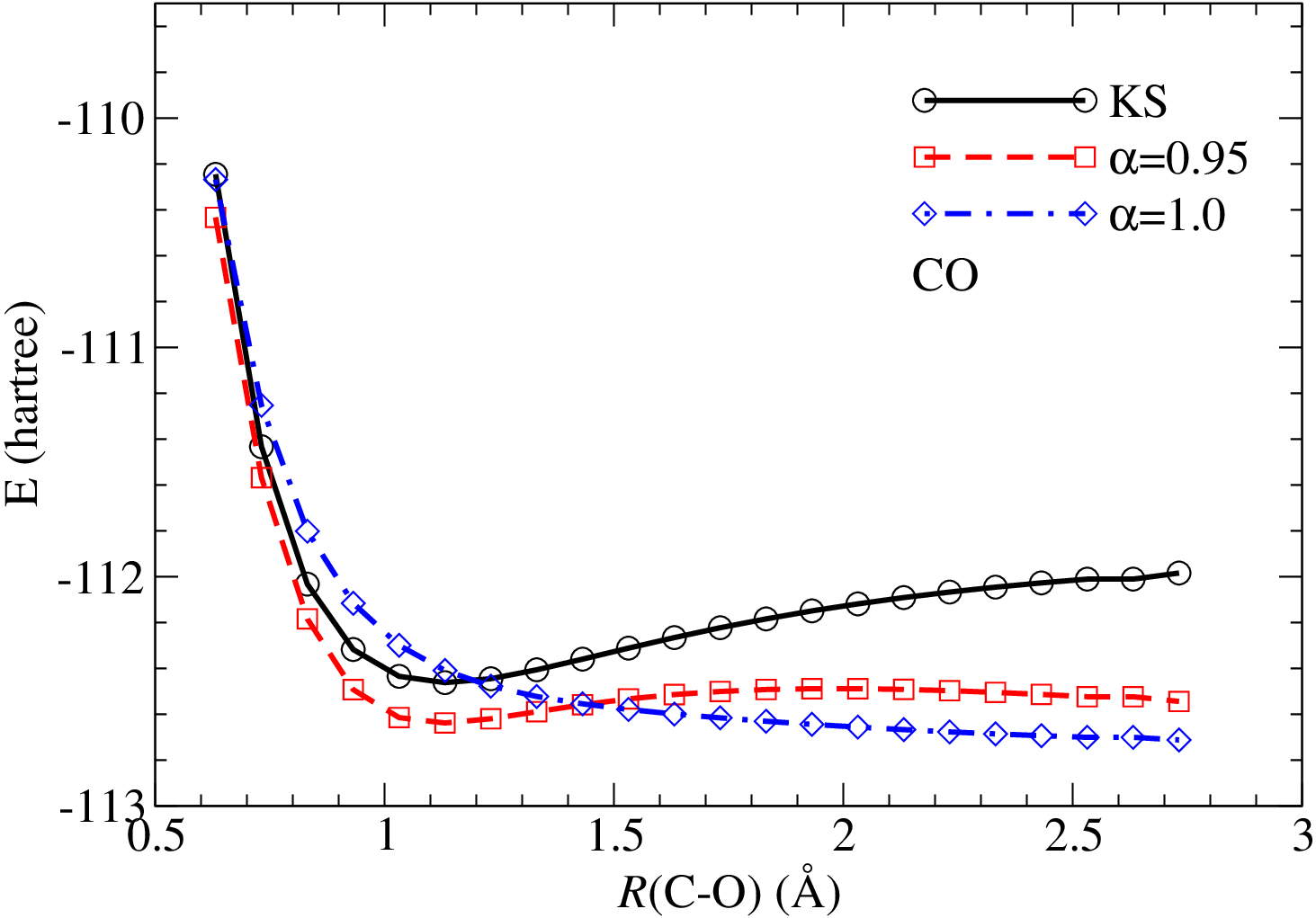}
\includegraphics*[width=8.0cm]{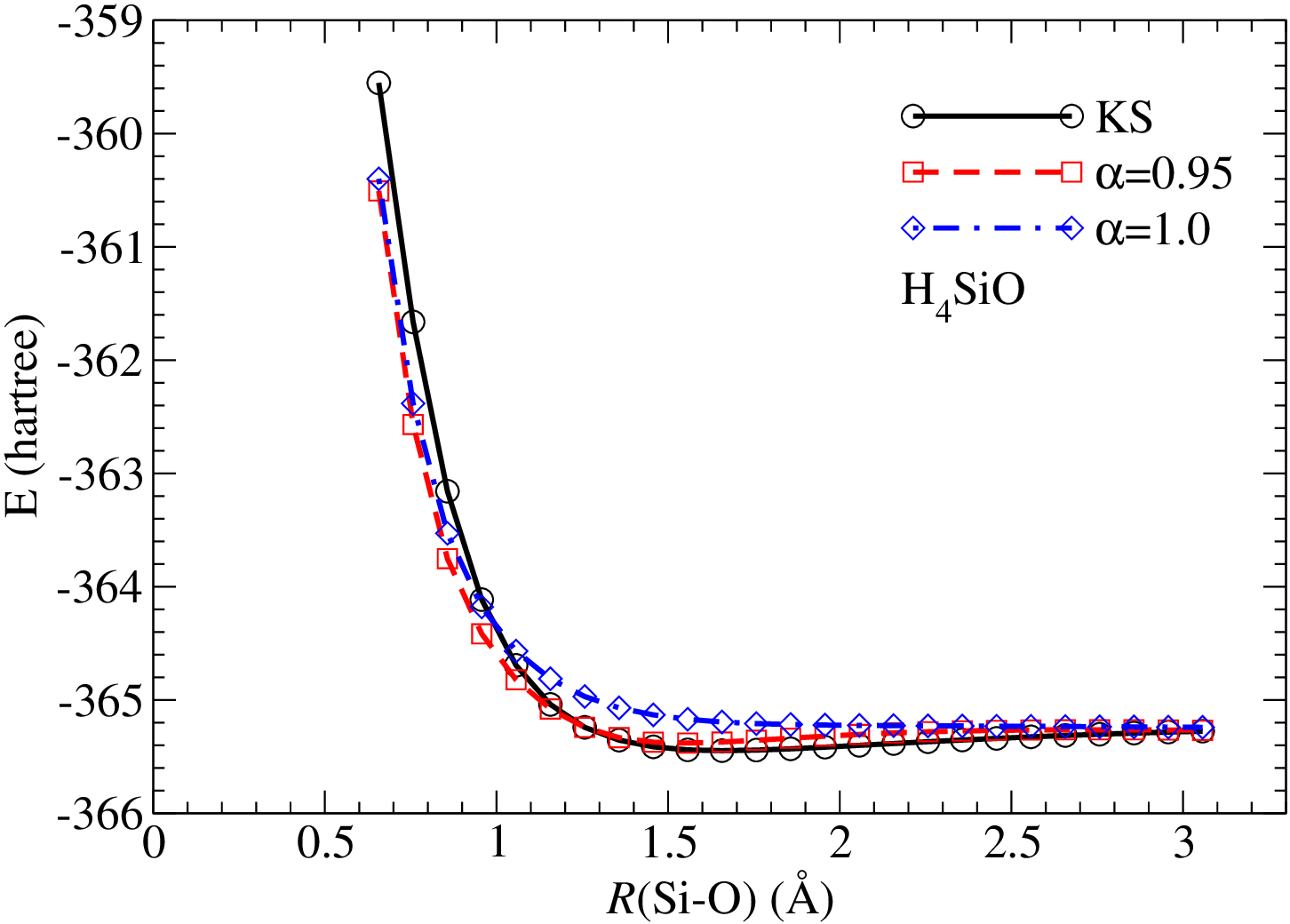}
\caption{
Total energy as a function of bond distance for the CO (left)
and H$_4$SiO$_4$ (right) molecules from the training set
obtained from a KS LDA calculation and from 
the post-KS orbital-free calculation with approximate GGA functionals.
}
\label{Etot-1}
\end{figure}

\begin{figure}
\includegraphics*[width=8.0cm]{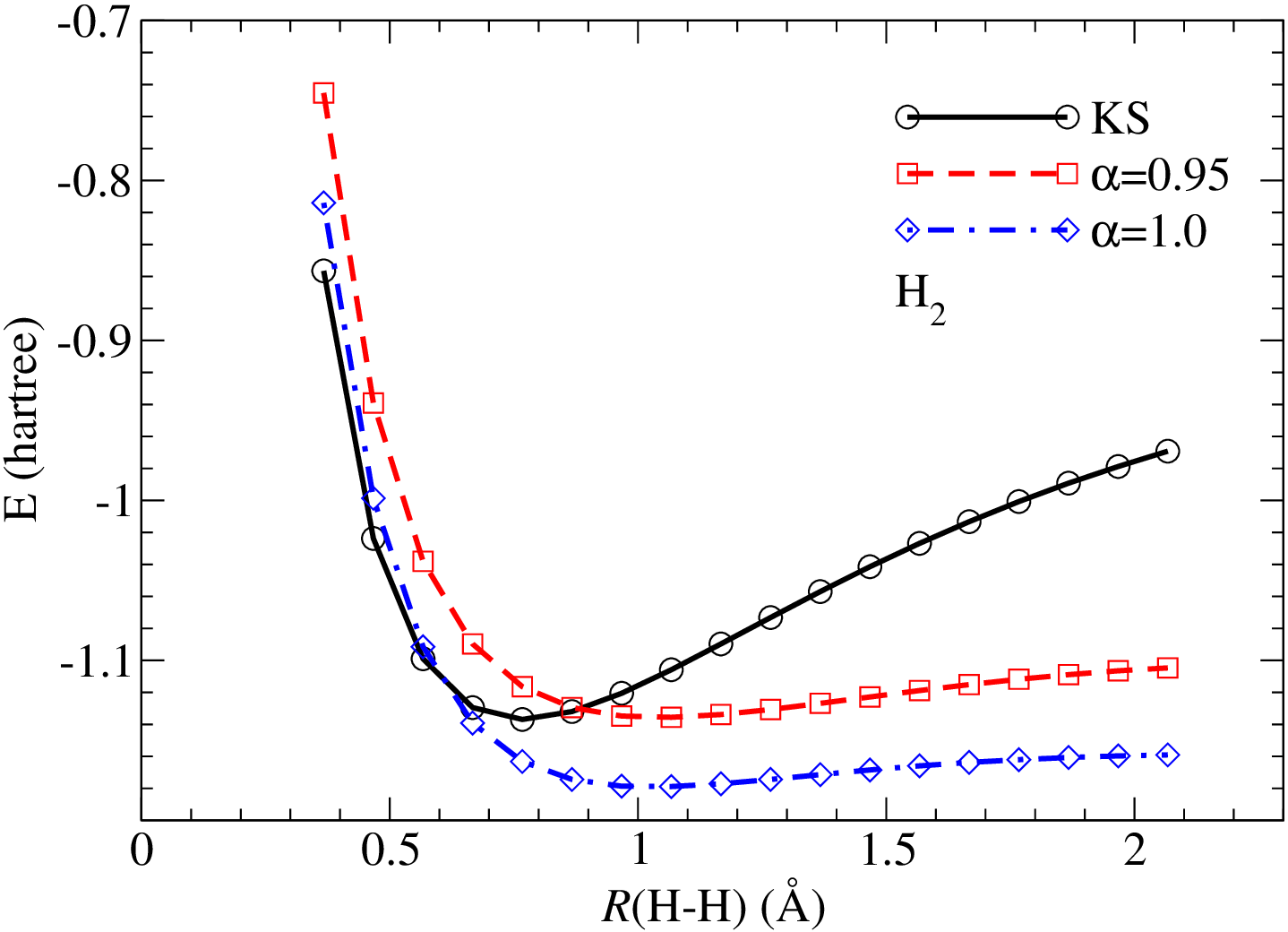}
\includegraphics*[width=8.0cm]{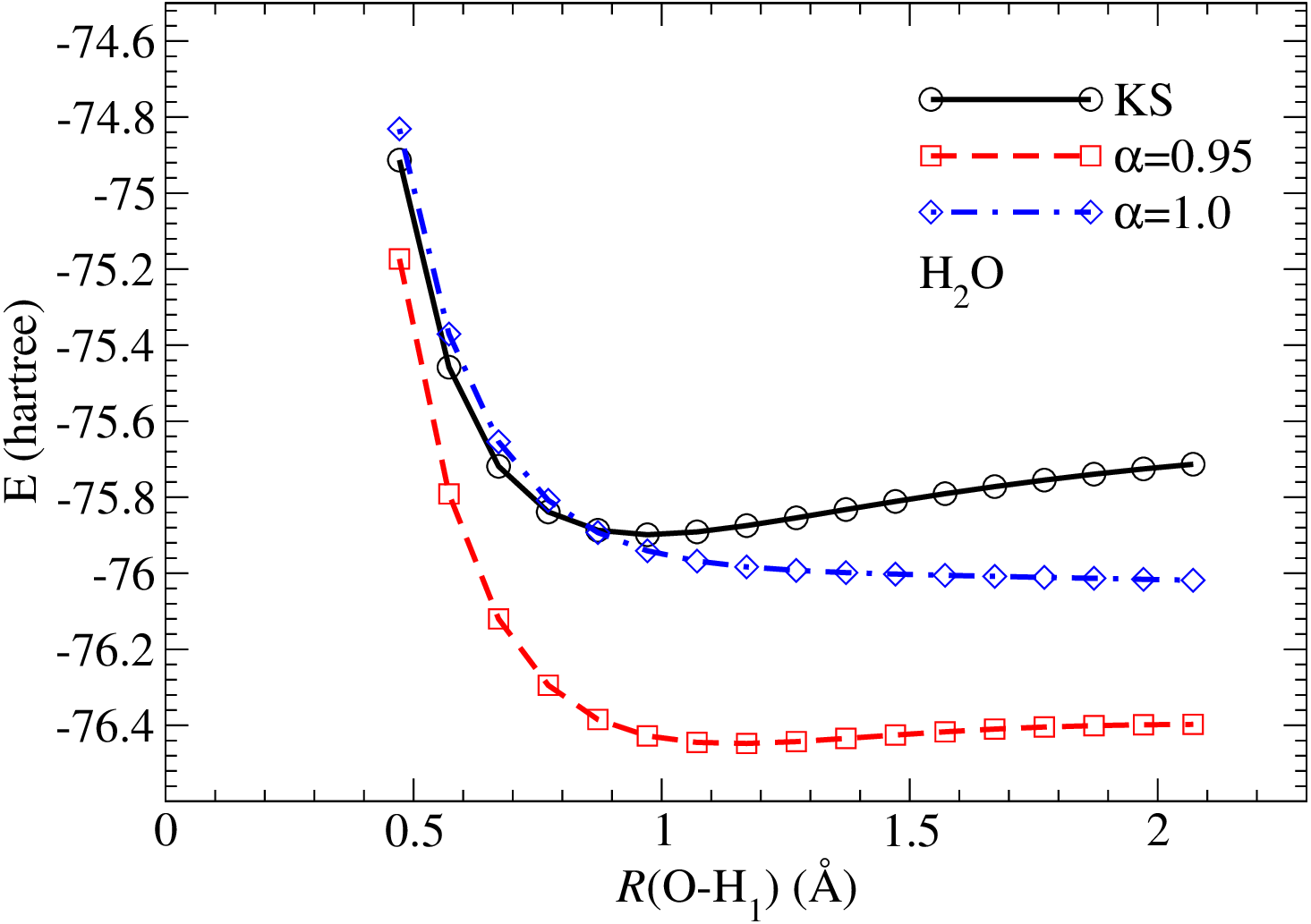}
\caption{
Total energy as a function of bond distance for the H$_2$ (left)
and H$_2$O (right) molecules (neither in training set)
obtained from a KS LDA calculation and from 
the post-KS orbital-free calculation with approximate GGA functionals.
}
\label{Etot-2}
\end{figure}

The quality of energy curves for the CO and H$_4$SiO molecules 
corresponding to the $\alpha=0.95$ functional is shown in Fig.\ \ref{Etot-1}.
The $\alpha=1$ curves (functional parameters fitted to optimize 
$\omega(\alpha=1)$, i.e.\ pure $E$-criterion) are shown for comparison. 
They have no minima. 
For the CO molecule, the minimum at $\alpha=0.95$ curve is too shallow 
compared to the reference KS result. In contrast, for H$_4$SiO the
agreement between the $\alpha=0.95$ 
optimized functional and KS results is excellent. Fig.\ \ref{Etot-2} 
shows a similar comparison for two molecules not in the training set. 
For the simplest, H$_2$, both functionals ($\alpha=0.95$ and $\alpha=1.0$)
provide qualitatively correct binding, but the minimum is too shallow
and there are large discrepancies at the dissociation limit 
relative to the KS result.
Single-bond stretching in H$_2$O is described qualitatively roughly correctly
only by the $\alpha=0.95$ functional.  As expected, the $\alpha = 1.0$ 
functional fails to give binding.

\begin{table}[htb]
\caption{\label{tab:table2}
Coefficients in Eq. (\ref{Ft})
from optimizing $\omega(\alpha=0.95)$.
}
\label{table:2}
\newcommand{\m}{\hphantom{$-$}}
\newcommand{\cc}[1]{\multicolumn{1}{c}{#1}}
\renewcommand{\tabcolsep}{2pc} 
\renewcommand{\arraystretch}{1.2} 
\begin{tabular}{@{}lr}
\hline
coefficient &  \rm value \\
\hline
$a_{ 1}=b_1$ &      12.100994770272     \\
$a_{ 2}=5/27+b_2$ & 10.829496969896\\
$a_{ 3}$ & -27.327919841144\\
$a_{ 4}$ &  73.841590552393\\
$a_{ 5}$ &  25.096089580269\\
$a_{ 6}$ & -45.306369888376\\
$a_{ 7}$ & -77.901835391837\\
$a_{ 8}$ & -20.862438996553\\
$a_{ 9}$ &  67.083330246208\\
\hline	
$b_{ 1}$ &  12.100994770272\\
$b_{ 2}$ &  10.644311784711\\
$b_{ 3}$ &  14.896876304511\\
$b_{ 4}$ &  5.5830951758904\\
$b_{ 5}$ & -24.558524755221\\
$b_{ 6}$ & -31.914940553009\\
$b_{ 7}$ &  4.3293607988211\\
$b_{ 8}$ &  17.169012815532\\
$b_{ 9}$ &  2.4210601059537\\
$b_{10}$ &  3.2527234245842\\
\hline
\end{tabular}\\[2pt]
\end{table}

Table \ref{tab:table2} lists the set of coefficients in 
the kinetic energy enhancement factor Eq. (\ref{Ft}) 
for the optimized $\omega(\alpha=0.95)$.

\section{Summary Discussion}

A GGA OFKE form is the simplest one-point functional which explicitly 
includes effects of electron density inhomogeneity.  
The GGA form, with parameters determined by constraints, has been very 
successful for approximate XC functionals, to the point that such   
functionals dominate in practical calculations.   A crucial distinction 
with respect to OFKE is 
the order of magnitude of the XC energy, about 10\% of the total energy.
The ground state KE, however, has the same order of magnitude
as the total energy and the KS KE is a large fraction of the total KE.
Hence a relative error of even a few 
percent in an approximate OFKE functional will have much bigger impact 
on calculated properties than would the same relative error in
an approximate XC functional.  

This distinction has significant implications for the constraint-based
development of GGA OFKE functionals. There is a related, but perhaps
less-obvious distinction.  Because exchange in physical systems is
purely Coulombic and correlation (as defined in DFT; recall
Eq.\ (\ref{Ecdefn})) is mostly Coulombic (the KE contribution is small),
it is eminently sensible to impose Coulombic constraints on an
approximate $E_{xc}$.  The resulting functional should be applicable
to a broad range of $v_{ext}$, if not truly universal.  Experience
shows that to be the case.  Good GGA $E_{xc}$ functionals deliver
variations in MARE over classes of properties and types of bonding but
they are broadly applicable.

Generating useful constraints on a GGA OFKE functional that preserve
universality has not been as straightforward so far as for the XC 
functionals.  One way to see the underlying difficulty is that $T_s$ is
a non-interacting system quantity, whereas $E_{xc}$ is, as just remarked,
predominantly a Coulombic quantity.  This means a lack of specificity 
for $T_s$ compared to $E_{xc}$.  Surmounting that lack is what
we have done by  abandoning universality and imposing conditions that
follow from Coulombic $v_{ext}$.  Precisely because they are non-universal
conditions, use of pseudo-potentials or Gaussian-type basis sets 
immediately introduces inconsistencies.  In Section 3 we have shown
how the consequences can be delineated clearly with simple one-center 
densities.

Section 4 then considered whether any GGA can have acceptable 
errors in both total energies and binding energies.  The result is 
not as encouraging as we would like.  For a small training set, the
best empirical OFKE GGA we have been able to develop so far 
provides a relatively small MARE for the total energy 
($\omega_E \approx 0.8\%$) but only semi-quantitative binding 
($\omega_{\Delta E} \approx 44\%$).
The use of mixed $E$-$\Delta E$ criteria is essential to get
both correct total energies and roughly reasonable binding simultaneously.

The correct second-order gradient expansion was the only
constraint imposed on the KE enhancement factor.  The result is
violation of a constraint, $F_\theta \ge 0$,
Eq.\ (\ref{FthetaPos}) which depends on a particular choice of
$t_\theta$.  We have not checked whether $v_\theta \ge 0$ is violated
for the empirical GGA, but are certain that it will be violated because
of the negative slope of $F_\theta^{empirical}$ for $s^2 \le 1$.  
One might hope that incorporation of
more constraints should make the functional better or, at least, 
more nearly universal in the
limited sense of improving its transferability to different systems 
and/or conditions.  The counterargument is that, except for the
high-order Pad\'e form itself, the empirical functional was
not restricted in any other way, a fact which should facilitate
optimization (even at the cost of realism or transferability).  

A weak caveat is that we have not yet tested the empirical functionals in 
SCF calculations.  This may be important, because the 
large-$s$ behavior of the empirical functionals 
for Pad\'e approximants of different orders might be
very different, while the $\omega_{E}$ and $\omega_{\Delta E}$ errors are 
very similar. 
That difference in 
large-$s$ behavior might be important for SCF calculations, but
not for post-KS calculations. However, the somewhat disappointing performance
on post-KS binding energy curves makes this, in our judgment, a 
somewhat problematic conjecture.  

During minimization of $\omega(\alpha)$ with respect to the 
independent parameters of the Pad{\'e}  enhancement factor (at 
fixed $\alpha$ of course), we 
encountered three main difficulties.   
First is the precision of numerical integration required (very
dense radial and angular meshes) to handle roughness in  
the kinetic energy enhancement factor.  Second is that, for any
given $\alpha$, the  optimization process very frequently sticks 
in local minima.  Optimizations therefore were run for multiple 
values of $\alpha$.  Those results were analyzed to find one or a few
superior parameter sets (based on values of $\omega_E$ and $\omega_{\Delta E}$).
Then those parameter sets were used as initial ones to start a repeat  
optimization for all $\alpha$. Eventually this ``by hand'' procedure
yielded optimal sets of parameters for every $\alpha$.  
Thirdly, the existence of multiple enhancement factors 
which deliver very similar values of the objective function $\omega(\alpha)$ 
despite being very different functions of $s$ makes the final functional
for each $\alpha$ sensitive to the choice of the training set.
Increasing the size of the training set might help to overcome that 
difficulty.  But we note again that our purpose here is not to attempt
a general, widely applicable empirical GGA for OFKE but simply to find 
the best for a modest selection of molecules.  Even with that narrow goal,
the outcome seems to be that there are significant limits on what
can be expected of a GGA OFKE functional.     

\section{Acknowledgments}
We thank the University of Florida High-Performance Computing Center for 
computational resources and technical support. We thank Keith Runge for
help with the title.  

\end{document}